\documentclass[prl,twocolumn,amssymb]{revtex4}
\usepackage{graphicx}
\usepackage{bm}
\usepackage{amsmath,amssymb,fancybox}

\begin{document}
\title{Finding resource states of
measurement-based quantum
computing is harder than quantum computing}
\author{Tomoyuki Morimae}
\email{morimae@gunma-u.ac.jp}
\affiliation{ASRLD Unit, Gunma University, 1-5-1 Tenjin-cho Kiryu-shi
Gunma-ken, 376-0052, Japan}

\begin{abstract}
Measurement-based quantum computing enables universal quantum computing
with only adaptive single-qubit measurements on certain many-qubit states, 
such as the graph state, the Affleck-Kennedy-Lieb-Tasaki (AKLT) state,
and several tensor-network states.
Finding new resource states of measurement-based quantum
computing is a hard task,
since for a given state there are exponentially
many possible measurement patterns on the state.
In this paper, we consider the problem of deciding,
for a given state and a set of unitary operators,
whether there exists a way of
measurement-based quantum computing on the state
that can realize all unitaries in the set, or not.
We show that the decision problem
is QCMA-hard, which means that 
finding new resource states of measurement-based quantum
computing is harder than quantum computing itself
(unless BQP is equal to QCMA).
We also derive an upperbound of the decision problem:
the problem is in a quantum version
of the second level of the polynomial hierarchy.
\end{abstract}
\maketitle  

Measurement-based quantum computing~\cite{MBQC} is another model of
quantum computing than the traditional circuit model
where universal quantum computing
can be done with only adaptive single-qubit measurements
on certain many-qubit states which are called
resource states. 
Although it is mathematically equivalent to the circuit model,
the clear separation between the resource preparation phase
and the resource consumption phase has enabled
plenty of new results in, for example, fault-tolerant
quantum computing~\cite{RHG}, 
condensed matter physics~\cite{Miyake,Miyake2,Miyake3,Wei,Else,Miller},
studying roles of quantumness in quantum 
computing~\cite{Gross,Bremner},
secure quantum computing (blind quantum computing)~\cite{BFK,HM},
and quantum complexity theory~\cite{Matt,MNS,FM}.

The first and the most standard example of 
universal resource states is the graphs state~\cite{MBQC},
which is obtained by applying $CZ$ operators
on all connected $|+\rangle$ states placed on
every vertex of a graph.
Researchers have tried to find more condensed-matter physically
motivated resource states.
For example, the Affleck-Kennedy-Lieb-Tasaki (AKLT)
state~\cite{AKLT} was found to be a universal 
resource state~\cite{Miyake,Miyake2,Miyake3,Wei}.
Several tensor-network states were also shown to be
universal resource states by considering virtual
quantum computing in the correlation space~\cite{Gross2}.
Furthermore, low-temperature thermal equilibrium states of
some physically motivated Hamiltonians were shown to
be universal resource states for topological measurement-based
quantum computing~\cite{Li,FM2}.
In spite of much efforts, however, 
we have only a very short list of universal resource states.
One of the main reasons of the difficulty of finding
new resource states is the
exponential increase of possible measurement patterns on 
a given state. 
Therefore we have a natural question: how hard is it
to find a new resource state? Is it, say, NP-hard?

In this paper, we study the computational complexity 
of finding new resource states of measurement-based quantum computing.
We consider the problem of deciding,
for a given state and a set of unitary operators,
whether there exists a way of
measurement-based quantum computing on the state
that can realize all unitaries in the set, or not.
We show that the decision problem is QCMA-hard. 
The class QCMA~\cite{QCMA,Aharonov} 
is a quantum version of NP and defined in the following way:

A language $L$ is in QCMA if and only if
there exists a uniformly-generated family
$\{V_x\}_x$ of polynomial-size quantum circuits
such that
\begin{itemize}
\item
If $x\in L$, then 
there exists a $w$-bit string $y\in\{0,1\}^w$ such that
the probability of obtaining 1 when the
first qubit of $V_x(|y\rangle\otimes|0^n\rangle)$ is measured
in the computational basis is $\ge \frac{2}{3}$.
Here, $n=poly(|x|)$
and $w=poly(|x|)$.
\item
If $x\notin L$, then 
for any $w$-bit string $y\in\{0,1\}^w$, 
the probability is $\le \frac{1}{3}$.
\end{itemize}
It is known that
the error bound $(\frac{2}{3},\frac{1}{3})$ can be amplified
to $(1-2^{-r},2^{-r})$ for any polynomial $r$
by using the standard argument of the error reduction
used in other probabilistic classes such as BPP, MA, and BQP.
Obviously QCMA contains BQP. (We have only to ignore
the witness.)
Moreover, QCMA seems to be strictly larger than BQP,
since it seems to be difficult 
to find a correct $y$ 
in a quantum polynomial
time. 
In fact, there are several results
that support ${\rm QCMA}\neq{\rm BQP}$.
(For example, it is obvious that QCMA contains NP. However, BQP
is not believed to contain NP~\cite{BBB+97}.)
Therefore, if we assume
${\rm QCMA}\neq{\rm BQP}$,
we can put our result concisely as follows:
``finding new resource states
is harder than quantum computing itself".

We also study upperbounds of the problem.
We show that the problem is in a quantum version of
the second level of the polynomial hierarchy.
The polynomial hierarchy is one of the most important concepts
in complexity theory, and its quantum versions were
considered in Refs.~\cite{Kampe,Yamakami}.

{\it Measurement-based quantum computing}.---
Before giving the precise definition of the problem that we show to
be QCMA-hard, let us here explain 
an abstract form of measurement-based quantum computing.
Assume that as a resource state, 
we are given an $N$-qubit state $|\Psi\rangle$.
Let 
$
{\mathcal U}\equiv\{U_y\}_{y\in\{0,1\}^w} 
$
be a set of unitary operators acting on $n$ qubits
$(n\le N)$.
We say that the resource state $|\Psi\rangle$ is ${\mathcal U}$-universal
with precision $\epsilon$
$(0\le\epsilon\le1)$
if 
there exists a polynomial-time classical algorithm
$\Lambda$ such that
for any $y\in\{0,1\}^w$,
\begin{itemize}
\item[1.]
We input $(1,y)$ to $\Lambda$. $\Lambda$ outputs
a classical description of a single-qubit unitary operator
$u_1$.
We measure the first qubit of the resource state in
the basis $\{u_1|0\rangle,u_1|1\rangle\}$.
We obtain the measurement result $m_1\in\{0,1\}$.
\item[2.]
We input $(2,y,m_1)$ to $\Lambda$. $\Lambda$ outputs
a classical description of a single-qubit unitary operator
$u_2$.
We measure the second qubit of the resource state in
the basis $\{u_2|0\rangle,u_2|1\rangle\}$.
We obtain the measurement result $m_2\in\{0,1\}$.
\item[3.]
We input $(3,y,m_1,m_2)$ to $\Lambda$. $\Lambda$ outputs
a classical description of a single-qubit unitary operator
$u_3$.
We measure the third qubit of the resource state in
the basis $\{u_3|0\rangle,u_3|1\rangle\}$.
We obtain the measurement result $m_3\in\{0,1\}$.
\item[4.]
In this way,
we repeat this adaptive single-qubit measurements
until all but the last $n$ qubits of the resource state
are measured.
In other words, when we measure $j$th qubit of
the resource state, we input $(j,y,m_1,...,m_{j-1})$
to $\Lambda$, and get a classical description of a single-qubit
unitary operator $u_j$ from $\Lambda$. We then
measure $j$th qubit
of the resource state in the basis $\{u_j|0\rangle,u_j|1\rangle\}$, 
and obtain the measurement result $m_j$.
We repeat it until $j=N-n$.
Let $|\psi_m'\rangle$,
where $m\equiv(m_1,...,m_{N-n})$,
be the (normalized) post-measurement state of
$n$ qubits of the resource state that are not measured.
We also denote the probability of obtaining $m$ by $p_m$.  
(For example, if $|\Psi\rangle$ is the graph state,
$p_m=2^{-(N-n)}$ for all $m$, and
$|\psi_m'\rangle$ is equal to $U_y|0^n\rangle$
up to some Pauli byproduct operators.)
\item[5.]
We input $(N-n+1,y,m)$ to $\Lambda$.
$\Lambda$ outputs classical descriptions
of single-qubit unitary operators 
$\{v_j\}_{j=1}^n$.
We apply $v_j$ on $j$th qubit of 
$|\psi_m'\rangle$
to obtain 
$
|\psi_m\rangle\equiv
\bigotimes_{j=1}^nv_j
|\psi_m'\rangle.
$
(This process is the ``final byproduct correction".
For example, if $|\Psi\rangle$ is the graph state,
each $v_j$ is a Pauli byproduct operator, and 
$|\psi_m\rangle=U_y|0^n\rangle$ for all $m$.)
\item[6.]
The state $\sum_mp_m|\psi_m\rangle\langle\psi_m|$
is close to the ideal state $U_y|0^n\rangle$
in the sense of
\begin{eqnarray*}
\frac{1}{2}\Big\|
\sum_m p_m |\psi_m\rangle\langle\psi_m|
-U_y|0^n\rangle\langle0^n|U_y^\dagger\Big\|_1\le \epsilon.
\end{eqnarray*}
Here, $\|X\|_1\equiv\mbox{Tr}\sqrt{X^\dagger X}$
is the trace norm.
\end{itemize}

{\it The problem}.---
Now we define the decision problem that we study,
which is a promise version of deciding whether a given 
state is non-${\mathcal U}$-universal or not.
We call the problem ${\rm NONUNIVERSALITY}_\epsilon$
for a parameter $\epsilon$ 
$(0\le\epsilon\le1)$:
\begin{itemize}
\item
Input: ${\mathcal U}$ and 
$|\Psi\rangle$.
\item
YES: $|\Psi\rangle$ is not ${\mathcal U}$-universal.
In other words, for any $\Lambda$ there exists $y$ such that,
\begin{eqnarray*}
\frac{1}{2}\Big\|
\sum_mp_m
|\psi_m\rangle\langle\psi_m|-
U_y|0^n\rangle\langle0^n|U_y^\dagger\Big\|_1\ge 1-\epsilon.
\end{eqnarray*}
\item
NO: $|\Psi\rangle$ is ${\mathcal U}$-universal.
In other words, there exists $\Lambda$
such that for all $y$
\begin{eqnarray*}
\frac{1}{2}\Big\|
\sum_mp_m
|\psi_m\rangle\langle\psi_m|-
U_y|0^n\rangle\langle0^n|U_y^\dagger\Big\|_1\le\epsilon.
\end{eqnarray*}
\end{itemize}
The main result of the present paper is
that the problem is QCMA-hard
for $\epsilon=2^{-t}$, where $t$ is any polynomial.

\begin{figure}[htbp]
\begin{center}
\includegraphics[width=0.3\textwidth]{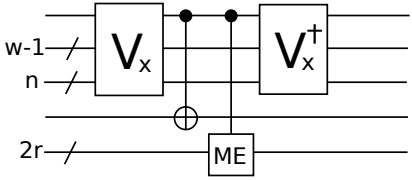}
\end{center}
\caption{
The unitary operator $U$.
ME means the unitary operator that changes
$|0^{2r}\rangle$  
to the maximally entangled state $|ME\rangle$.
} 
\label{fig}
\end{figure}

{\it Proof}.---
Here we give a proof.
Let us assume that a language $L$ is in ${\rm QCMA}$,
and let $V_x$ be the corresponding verification
circuit for an instance $x$.
Due to the standard argument of
the error reduction, we can assume without loss
of generality that
the acceptance probability $p$ satisfies $p\ge 1-2^{-r}$
if $x\in L$ and $p\le 2^{-r}$ if $x\notin L$,
where $r$ is any polynomial.
Fix one $x$.
From $V_x$, we construct the unitary operator $U$ 
that acts on $n+w+2r+1$ qubits as follows
(see Fig.~\ref{fig}):
\begin{itemize}
\item[1.]
Apply $V_x$ on $|y0^n\rangle$ to generate $V_x|y0^n\rangle$.
\item[2.]
Add an ancilla qubit initialized in $|0\rangle_1$ to generate
$V_x|y0^n\rangle\otimes|0\rangle_1$.
\item[3.]
Flip the ancilla qubit if and only if the first qubit
of $V_x|y0^n\rangle$ is $|1\rangle$.
We therefore obtain
\begin{eqnarray*}
&&(|0\rangle\langle0|\otimes I^{\otimes n+w-1})
V_x|y0^n\rangle|0\rangle_1\\
&+&(|1\rangle\langle1|\otimes I^{\otimes n+w-1})
V_x|y0^n\rangle|1\rangle_1.
\end{eqnarray*}
Here $I\equiv|0\rangle\langle0|+|1\rangle\langle1|$
is the two-dimensional identity operator.
\item[4.]
Add $2r$ ancilla qubits initialized in $|0^{2r}\rangle_{2}$.
We obtain
\begin{eqnarray*}
&&(|0\rangle\langle0|\otimes I^{\otimes n+w-1})
V_x|y0^n\rangle|0\rangle_1|0^{2r}\rangle_2\\
&+&(|1\rangle\langle1|\otimes I^{\otimes n+w-1})
V_x|y0^n\rangle|1\rangle_1|0^{2r}\rangle_2.
\end{eqnarray*}
\item[5.]
Apply a $2r$-qubit unitary operator $ME$ on the
$2r$ ancilla qubits $|0^{2r}\rangle_2$
that changes the state $|0^{2r}\rangle$ to the
maximally-entangled state
\begin{eqnarray*}
|ME\rangle\equiv\frac{1}{\sqrt{2^r}}
\sum_{j=1}^{2^r}|j\rangle|j\rangle
\end{eqnarray*}
if and only the first qubit of $V_x|y0^n\rangle$ is $|1\rangle$.
We therefore obtain
\begin{eqnarray*}
&&(|0\rangle\langle0|\otimes I^{\otimes n+w-1})
V_x|y0^n\rangle|0\rangle_1|0^{2r}\rangle_2\\
&+&(|1\rangle\langle1|\otimes I^{\otimes n+w-1})
V_x|y0^n\rangle|1\rangle_1|ME\rangle_2.
\end{eqnarray*}
\item[6.]
Apply $V_x^\dagger$ on the main register.
We thus obtain
the final state
\begin{eqnarray*}
&&U|y0^n\rangle|0\rangle_1|0^{2r}\rangle_2\\
&=&V_x^\dagger(|0\rangle\langle0|\otimes I^{\otimes n+w-1})
V_x|y0^n\rangle|0\rangle_1|0^{2r}\rangle_2\\
&+&V_x^\dagger(|1\rangle\langle1|\otimes I^{\otimes n+w-1})
V_x|y0^n\rangle|1\rangle_1|ME\rangle_2.
\end{eqnarray*}
\end{itemize}

Let us define 
\begin{eqnarray*}
U_y\equiv
U\Big[\Big(\bigotimes_{j=1}^w X_j^{y_j}\Big)
\otimes I^{\otimes n+2r+1}\Big],
\end{eqnarray*}
where $y_j$ is the $j$th bit of $y$,
and $X_j\equiv|0\rangle\langle1|+|1\rangle\langle0|$
is the bit-flip operator acting on $j$th qubit.
We also define
\begin{eqnarray*}
|\Psi\rangle\equiv
|0^{n+w+2r+2}\rangle.
\end{eqnarray*}

First, we consider the case of $x\in L$.
In this case, $p\ge1-2^{-r}$ for a certain $y$.
Note that whatever $\Lambda$ we choose,
what we can do is just measuring a single qubit of
$|\Psi\rangle$ and then rotating each of
the $n+w+2r+1$ unmeasured qubits.
Therefore, for any $\Lambda$, $|\psi_m\rangle$
is an $(n+w+2r+1)$-qubit product state
for all $m$.
Therefore, from Uhlmann's theorem,
\begin{eqnarray*}
&&|\langle \psi_m|U_y|0^{n+w+2r+1}\rangle|^2\\
&=&
|\langle \psi_m|U|y0^{n+2r+1}\rangle|^2\\
&\le&
F(\rho,\sigma)^2\\
&=&(1-p)|\langle\xi_1\xi_2|0^{2r}\rangle|^2
+p|\langle\xi_1\xi_2|ME\rangle|^2\\
&\le&(1-p)+p|\langle\xi_1\xi_2|ME\rangle|^2\\
&\le&(1-p)+pF\Big(|\xi_1\rangle\langle\xi_1|,
\frac{I^{\otimes r}}{2^r}\Big)^2\\
&=&1-p+\frac{p}{2^r}\\
&\le&2^{-r}+2^{-r}\\
&=&2^{-r+1},
\end{eqnarray*}
where
$
F(\rho,\sigma)\equiv\mbox{Tr}\sqrt{\sqrt{\rho}\sigma\sqrt{\rho}}
$
is the fidelity,
\begin{eqnarray*}
\rho&\equiv&
\mbox{Tr}_2(|\psi_m\rangle\langle\psi_m|)\\
&\equiv&|\xi_1\rangle\langle\xi_1|\otimes
|\xi_2\rangle\langle\xi_2|,\\
\sigma&\equiv&
\mbox{Tr}_2(U|y0^{n+2r+1}\rangle\langle y0^{n+2r+1}|U^\dagger)\\
&=&(1-p)|0^{2r}\rangle\langle0^{2r}|+p|ME\rangle\langle ME|,
\end{eqnarray*}
$\mbox{Tr}_2$ is the partial trace except for the send
ancilla register, and 
$|\xi_j\rangle$ $(j=1,2)$ is a certain (actually product)
$r$-qubit state.
Therefore
\begin{eqnarray*}
&&F\Big(\sum_mp_m|\psi_m\rangle\langle\psi_m|,
U_y|0^{n+w+2r+1}\rangle\langle0^{n+w+2r+1}|U_y^\dagger\Big)\\
&=&\sqrt{\sum_mp_m|\langle\psi_m|U_y|0^{n+w+2r+1}\rangle|^2}\\
&\le&2^{\frac{-r+1}{2}}.
\end{eqnarray*}
Hence we have shown that for any $\Lambda$, there exists $y$
such that
\begin{eqnarray*}
&&\frac{1}{2}\Big\|
\sum_mp_m|\psi_m\rangle\langle\psi_m|
-U_y|0^{n+w+2r+1}\rangle\langle0^{n+w+2r+1}|U_y^\dagger
\Big\|_1\\
&\ge& 1-2^{\frac{-r+1}{2}}\\
&\ge& 1-2^{-t},
\end{eqnarray*}
where we have taken $r\ge 2t+1$.

Next, we consider the case of $x\notin L$.
In this case, $p\le 2^{-r}$ for any $y$.
We define $\Lambda$ in such a way that
\begin{eqnarray*}
|\psi_m\rangle=|y0^{n+2r+1}\rangle
\end{eqnarray*}
for any $m$.
This is trivially possible as follows: 
\begin{itemize}
\item[1.]
Measure the first qubit
of $|\Psi\rangle=|0^{n+w+2r+2}\rangle$ in the computational basis.
Then we obtain $|\tilde{\psi}_{m=0}\rangle=|0^{n+w+2r+1}\rangle$
with probability 1.
\item[2.]
Apply $X_j^{y_j}$ on the $j$th qubit of
$|\tilde{\psi}_{m=0}\rangle$ for $j=1,...,w$ to obtain
$|\psi_{m=0}\rangle=|y0^{n+2r+1}\rangle$.
\end{itemize}
Then, for any $y$,
\begin{eqnarray*}
&&\sum_mp_m|\langle\psi_m|U_y|0^{n+w+2r+1}\rangle|^2\\
&=&|\langle\psi_{m=0}|U_y|0^{n+w+2r+1}\rangle|^2\\
&=&|\langle y0^{n+2r+1}|U|y0^{n+2r+1}\rangle|^2\\
&=&
|
\langle y0^n|V_x^\dagger
(|0\rangle\langle0|\otimes I^{\otimes n+w-1})
V_x|y0^n\rangle|^2\\
&=&(1-p)^2\\
&\ge&
(1-2^{-r})^2\\
&=&1-2^{-r+1}+2^{-2r}\\
&\ge&1-2^{-r+1}.
\end{eqnarray*}
Therefore
\begin{eqnarray*}
&&F\Big(\sum_mp_m|\psi_m\rangle\langle\psi_m|,
U_y|0^{n+w+2r+1}\rangle\langle0^{n+w+2r+1}|U_y^\dagger\Big)\\
&=&\sqrt{\sum_mp_m|\langle\psi_m|U_y|0^{n+w+2r+1}\rangle|^2}\\
&\ge&\sqrt{1-2^{-r+1}}.
\end{eqnarray*}
Hence we have shown that there exists $\Lambda$ such that
for any $y$
\begin{eqnarray*}
&&\frac{1}{2}\Big\|
\sum_mp_m|\psi_m\rangle\langle\psi_m|
-U_y|0^{n+w+2r+1}\rangle\langle0^{n+w+2r+1}|U_y^\dagger
\Big\|_1\\
&\le& 2^{\frac{-r+1}{2}}\\
&\le& 2^{-t}.
\end{eqnarray*}
In summary, we have shown that the promise problem is
QCMA-hard.

{\it Upperbound}.---
In this paper, we have shown that the problem
NONUNIVERSALITY is QCMA-hard.
In other words, we have derived an lower bound of the
problem.
It is an important open problem to find any better lower bound
and better upper bound of the problem, or to show that the problem
is complete for a complexity class.
Here we point out that a quantum version of
$\Pi_2$, which we call ${\rm Q}\Pi_2$, 
is an upper bound of the problem.
We define the class ${\rm Q}\Pi_2$ as follows:

A language $L$ is in ${\rm Q}\Pi_2(a,b)$ if and only if
there exists a uniformly generated family $\{V_x\}_x$ of
polynomial-size quantum circuits such that
\begin{itemize}
\item
If $x\in L$ then for any $\lambda$-bit string $\Lambda$
there exists a $w$-bit string $y$ such that
the probability of obtaining 1 when the first
qubit of $V_x(|\Lambda\rangle|y\rangle|0^n\rangle)$
is measured in the computational basis is $\ge a$.
Here, $\lambda,w,n=poly(|x|)$.
\item
If $x\notin L$ then
there exists a $\lambda$-bit string $\Lambda$
such that for any $w$-bit string $y$ 
the probability 
is $\le b$.
\end{itemize}
It is obvious that ${\rm Q}\Pi_2$ is in PSPACE.
(We have only to try all possible $\Lambda$ and $y$.)
Other types of quantum generalizations of the polynomial hierarchy were
studied in Refs.~\cite{Kampe,Yamakami}.

We can show that the problem NONUNIVERSALITY is
in ${\rm Q}\Pi_2(1-2\epsilon,2\epsilon)$.
In fact, since measurement-based quantum computing
can be simulated by a circuit model, there exists
a polynomial-size quantum circuit $V$ 
and polynomials $t$ and $n$ such that
the reduced density operator
of some $n$ qubits of
$V(|\Lambda\rangle|y\rangle|0^t\rangle)$
is
\begin{eqnarray*}
\rho\equiv
U_y^\dagger\Big(\sum_mp_m|\psi_m\rangle\langle\psi_m|\Big)U_y.
\end{eqnarray*}
We then measure all qubits of
$\rho$ in the computational basis, and reject
if and only if all qubits are 0.
The acceptance probability
$p$ is
therefore
\begin{eqnarray*}
p=
1-\langle 0^n|
U_y^\dagger
\Big(\sum_mp_m|\psi_m\rangle\langle\psi_m|\Big)U_y|0^n\rangle,
\end{eqnarray*}
which means
\begin{eqnarray*}
1-\sqrt{1-p}\le\frac{1}{2}\Big\|
\sum_mp_m|\psi_m\rangle\langle\psi_m|-
U_y|0^n\rangle\langle0^n|U_y^\dagger\Big\|_1
\le\sqrt{p}.
\end{eqnarray*}
Therefore, 
for the the yes case,
for any $\Lambda$, there exists $y$ such that
$1-\epsilon\le\sqrt{p}$, which means
\begin{eqnarray*}
p&\ge& 1-2\epsilon+\epsilon^2\\
&\ge&1-2\epsilon.
\end{eqnarray*}
For the no case,
there exists $\Lambda$ such that for any $y$,
$1-\sqrt{1-p}\le \epsilon$, which means
\begin{eqnarray*}
p&\le& 2\epsilon-\epsilon^2\\
&\le&2\epsilon.
\end{eqnarray*}
Hence we have shown that the problem is in 
${\rm Q}\Pi_2(1-2\epsilon,2\epsilon)$.

The author thanks Bill Fefferman,
Keisuke Fujii, 
Cedric Yen-Yu Lin,
and Harumichi Nishimura
for discussion.
The author is supported by
Grant-in-Aid for Scientific Research on Innovative Areas
No.15H00850 of MEXT Japan, and the Grant-in-Aid
for Young Scientists (B) No.26730003 of JSPS.


\begin{thebibliography}{00}
\bibitem{MBQC}
R. Raussendorf and H. J. Briegel,
A one-way quantum computer.
Phys. Rev. Lett. {\bf86}, 5188 (2001).
\bibitem{RHG}
R. Raussendorf, J. Harrington, and K. Goyal,
Topological fault-tolerance in cluster state quantum
computation.
New J. Phys. {\bf9}, 199 (2007).
\bibitem{Miyake}
G. K. Brennen and A. Miyake,
Measurement-based quantum computer in the gapped ground state
of a two-body Hamiltonian.
Phys. Rev. Lett. {\bf101}, 010502 (2008).
\bibitem{Miyake2}
A. Miyake,
Quantum computational capability of a 2D valence bond
solid phase.
Ann. Phys. {\bf326}, 1656 (2011).
\bibitem{Miyake3}
A. Miyake,
Quantum computation on the edge of a symmetry-protected
topological order.
Phys. Rev. Lett. {\bf105}, 040501 (2010).
\bibitem{Wei}
T. C. Wei, I. Affleck, and R. Raussendorf,
Affleck-Kennedy-Lieb-Tasaki state on a honeycomb
lattice is a universal quantum computational resource.
Phys. Rev. Lett. {\bf106}, 070501 (2011).
\bibitem{Else}
D. V. Else, I. Schwarz, S. D. Bartlett, and A. C. Doherty,
Symmetry-protected phases for measurement-based quantum
computation.
Phys. Rev. Lett. {\bf108}, 240505 (2012).
\bibitem{Miller}
J. Miller and A. Miyake,
Resource quality of a symmetry-protected topologically ordered
phase for quantum computation.
Phys. Rev. Lett. {\bf114}, 120506 (2015).
\bibitem{Gross}
D. Gross, S. T. Flammia, and J. Eisert,
Most quantum states are too entangled to be useful
as computational resources.
Phys. Rev. Lett. {\bf102}, 190501 (2009).
\bibitem{Bremner}
M. J. Bremner, C. Mora, and A. Winter,
Are random pure states useful for quantum computation?
Phys. Rev. Lett. {\bf102}, 190502 (2009).
\bibitem{BFK}
A. Broadbent, J. Fitzsimons, and E. Kashefi,
Universal blind quantum computation.
In Proceedings of the 50th Annual IEEE Symposium
on Foundations of Computer Science
(IEEE Computer Society, Los Alamitos, USA, 2009),
p.517.
\bibitem{HM}
M. Hayashi and T. Morimae,
Verifiable measurement-only blind quantum computing with
stabilizer testing.
Phys. Rev. Lett. {\bf115}, 220502 (2015).
\bibitem{Matt}
M. McKague,
Interactive proofs for BQP via self-tested graph
states.
Theor. of Comput. {\bf12}, 1 (2016).
\bibitem{MNS}
T. Morimae, D. Nagaj, and N. Schuch,
Quantum proofs can be verified using only single qubit
measurements.
Phys. Rev. A {\bf93}, 022326 (2016).
\bibitem{FM}
K. Fujii and T. Morimae, 
Quantum commuting circuits and complexity of Ising partition
functions.
arXiv:1311.2128
\bibitem{AKLT}
I. Affleck, T. Kennedy, E. H. Lieb, and H. Tasaki,
Valence bond ground states in isotropic quantum antiferromagnets.
Comm. Math. Phys. {\bf115}, 477 (1988).
\bibitem{Gross2}
D. Gross, J. Eisert, N. Schuch, and D. Perez-Garcia,
Measurement-based quantum computation beyond the
one-way model.
Phys. Rev. A {\bf76}, 052315 (2007).
\bibitem{Li}
Y. Li, D. E. Browne, L. C. Kwek, R. Raussendorf,
and T. C. Wei,
Thermal states as universal resources for quantum
computation with always-on interactions.
Phys. Rev. Lett. {\bf 107}, 060501 (2011).
\bibitem{FM2}
K. Fujii and T. Morimae,
Topologically protected measurement-based quantum computation
on the thermal state of a nearest-neighbor two-body
Hamiltonian with spin-3/2 particles.
Phys. Rev. A {\bf 85}, 010304(R) (2012).
\bibitem{QCMA}
S. Aaronson and G. Kuperberg,
Quantum versus classical proofs and advice.
Theor. of Comput. {\bf3}, 129 (2007).
\bibitem{Aharonov}
D. Aharonov and T. Naveh,
Quantum NP - a survey.
arXiv:0210077
\bibitem{BBB+97}
C. H. Bennett, E. Bernstein, G. Brassard,
and U. Vazirani,
Strength and weakness of quantum computing.
SIAM J. of Computing {\bf26}, 1510 (1997).
\bibitem{Kampe}
S. Gharibian and J. Kampe,
Hardness of approximation for quantum problems.
Quant. Inf. Comput. {\bf14}, 517 (2014).
\bibitem{Yamakami}
T. Yamakami,
Quantum NP and a quantum hierarchy.
In Proc. of the 2nd IFIP International Conference on Theoretical
Computer Science, pp.323-336. Kluwer Academic Publishers (2002).
\end{thebibliography}
\end{document}